# Antisymmetric magnetoresistance due to domain wall tilting in perpendicular magnetized films

Yangtao Su[1,2], Yang Meng[1,2]*, Haibin Shi[1,2], Li Wang[1,2], Xinyu Cao[1,2], Ying Zhang[1,2], Runwei Li[3], and Hongwu Zhao[1,2,4†]

[1]Beijing National Laboratory for Condensed Matter Physics, Institute of Physics, Chinese Academy of Sciences, Beijing 100190, China

[2]School of Physical Sciences, University of Chinese Academy of Sciences, Beijing 100049, China

[3]CAS Key Laboratory of Magnetic Materials and Devices, Ningbo Institute of Materials Technology and Engineering, Chinese Academy of Sciences, Ningbo, Zhejiang 315201, China

[4]Songshan Lake Materials Laboratory, Dongguan, Guangdong 523808, China

\* Author to whom correspondence should be addressed. ymeng@iphy.ac.cn

† Author to whom correspondence should be addressed. hwzhao@iphy.ac.cn

## ABSTRACT

We report the observation of the antisymmetric magnetoresistance (MR) in perpendicular magnetized CoTb films with inhomogeneous magnetization distribution driven by gradient magnetic field. By synchronously charactering the domain pattern evolution during transport measurements, we demonstrate that the nonequilibrium currents in the vicinity of tilting domain walls give rise to such anomalous MR. Moreover, theoretical calculation and analysis reveal that the geometry factor of the multidomain texture plays a dominant role in generating the nonequilibrium current. The explicitly established interplay between the anomalous transport behaviors and the particular domain wall geometry is essential to deepening understanding of the antisymmetric MR, and pave a new way for designing novel domain wall electronic devices.

In a Stoner ferromagnet, when the magnetization is homogeneous throughout the material, the magnetoresistance (MR) effect intrinsically originates from the spin orbit interaction between the spin polarized current and the magnetic moments [1,2]. Further studies on magnetic heterostructures with controllable magnetization have boosted intriguing and application-oriented discoveries of giant magnetoresistance and tunneling magnetoresistance effects [3-6]. However, when there are inhomogeneities in the magnetization direction, i.e., domain walls, distinct transport processes occur correspondingly, giving rise to various extrinsic anomalies MR [7-11]. Recently, experimental studies have demonstrated the presence of a single domain wall leads to novel antisymmetric magnetoresistance in magnetic films with perpendicular anisotropy [12-14]. This phenomenon has been attributed to the extraordinary Hall effect (EHE) induced by circulating currents in the vicinity of the single wall [15-18], where the domain wall, the current and the magnetization are mutually perpendicular. However, this "single wall" model is not applicable with respect to the special criteria to subsequently emerging antisymmetric MR in multidomain structures and in-plane magnetized films [17,19-21]. Indeed, experimental studies on clearly establishing the interplay between the anomalous transport behaviors and the particular magnetic domain texture can better help to clarify the origin of the antisymmetric anomaly.

In this letter, we demonstrate that the highly tunable domain wall tilting can result in antisymmetric MR in perpendicular magnetized CoTb films. By applying a controllable magnetic field gradient, a spatially inhomogeneous magnetization distribution has been implemented in the sample, leading to the magnetoresistance anomaly. Through synchronously measuring the magnetic domain structure and transport properties, we clearly show that such anomalous MR is mainly caused by the nonequilibrium current in the vicinity of the tilting domain walls of the multidomain structure. Theoretical calculation and analysis further indicate that the generated nonequilibrium current is basically determined by the geometry factor of the titling domain walls. Our findings are essential for quantitatively understanding the antisymmetric MR effect.

Ferrimagnetic $Co_xTb_{1-x}$ alloys exhibit excellent perpendicular magnetic anisotropy (PMA) and their magnetization reversal is primarily controlled by domain wall propagation[22,23]. Herein, we use $Co_{0.85}Tb_{0.15}$ as a model material to investigate the effect of inhomogeneous magnetization texture on its transport properties. The Pt(5 nm)/$Co_{0.85}Tb_{0.15}$(12 nm)/Pt(5 nm) films were deposited on Si substrates by magnetron sputtering. The $Co_{0.85}Tb_{0.15}$ film was prepared by co-sputtering Co and Tb targets with adjustable sputtering powers. During film deposition, a metallic mask was used to fabricate the Hall bar structure (width: 5 mm, length: 7 mm, arm width: 0.5 mm). The magnetic properties of $Co_{0.85}Tb_{0.15}$ films were investigated by the magneto-optic Kerr effect (MOKE). The transport measurements were carried out in a Physical Property Measurement System (PPMS-9) and a Janis ST-300 cryostat, respectively, where the magnetoresistance ( MR $= \frac{\Delta R}{R_{xx}(H=0)} = \frac{R_{xx}(H)-R_{xx}(H=0)}{R_{xx}(H=0)}$) and transverse Hall resistance $R_H$ were simultaneously measured. Using an EVICO Kerr microscopy, the magnetic domain structures were simultaneously characterized during magnetic transport measurements. The magnetic field $H$ was always applied normal to the film.

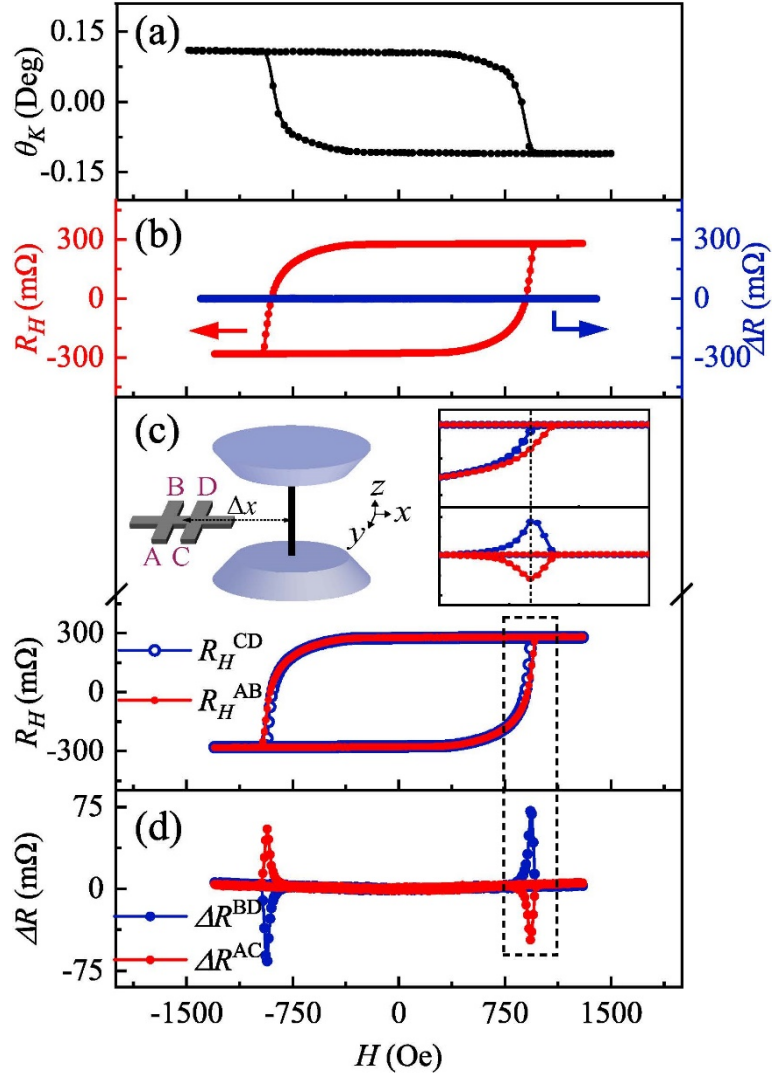

Figure 1 (a) MOKE hysteresis loop of Pt(5 nm)/Co$_{0.85}$Tb$_{0.15}$(12 nm)/Pt(5 nm). (b) Hall resistance $R_H$ and magnetoresistance change $\Delta R$ in uniform magnetic field. (c) $R_H$–$H$ curves in nonuniform field. Left inset: schematic of magnetoresistance measurement configuration. The offset distance $\Delta x = -13$ mm. Right inset: enlarged views of $R_H$–$H$ and $\Delta R$–$H$ curves. (d) Asymmetric $\Delta R$ in nonuniform field.

First, we investigated the magnetic and transport properties of the Co$_{0.85}$Tb$_{0.15}$ Hall bar under *uniform* magnetic fields. As shown in Fig. 1(a), the MOKE hysteresis loop shows sharp magnetization reversals with coercive field $H_c$ around 900 Oe, indicating the strong PMA of Co$_{0.85}$Tb$_{0.15}$[24]. The magnetic transport properties of the $Co_{0.85}Tb_{0.15}$ Hall bar were measured with a PPMS (magnetic field homogeneity < 0.01% over 5.5 cm on-axis). Compared with the *M-H* loop in Fig. 1(a), the Hall resistance $R_H$ in Fig. 1(b) shows the similar square shape and the same $Hc$, revealing that the $R_H$ is proportional to the magnetization along the field direction. Meanwhile, the simultaneously measured magnetoresistance change $\Delta R$ is almost near-zero, which can be attributed to the current-perpendicular-to-magnetization configuration inside magnetic domains and the negligible domain wall magnetoresistance [10,25].

Next, we examined the influence of inhomogeneous magnetization on the transport properties of $Co_{0.85}Tb_{0.15}$. As shown in the insert of Fig. 1(c), our custom-made electromagnets yield a heterogeneous field with homogeneity below 500 ppm within a cylindrical region of 5 mm in radius. Purposely, the $Co_{0.85}Tb_{0.15}$ Hall bar was fixed at an offset position $\Delta x = -13$ mm, the distance between the centers of magnets and the Hall bar, far away from the region of homogeneous field. As a result, the magnetic field applied at the AB junction of the Hall bar is actually smaller than that of the CD junction. The inhomogeneous gradient field directly leads to the slight difference between $R_H$-H curves measured at A-B and C-D electrode pairs around the saturation field (see the insert of Fig. 1(c)). The inconsistency clearly indicates the occurrence of asynchronous magnetization reversals in two Hall junctions. On the other hand, the simultaneously measured magnetoresistance $\Delta R$ across electrodes A and C exhibits obvious antisymmetric behavior (Fig. 1(d)), entirely different from the near-zero $\Delta R$ obtained in uniform field (Fig. 1(b)). The antisymmetric $\Delta R - H$ curves display a negative peak at 930 Oe for the ascending field branch, and a positive peak at -930 Oe for the descending field branch. By contrast, the $\Delta R$ measured across electrodes B and D displays similar antisymmetry but of opposite polarity (Fig. 1(d)). By carefully comparing $\Delta R$ and $R_H$ (see the enlarged views in Fig. 1(c)), it is clearly observed that when $\Delta R$ reaches its maximum, the magnetization of junction CD is nearly saturated and the magnetization of junction AB is not yet. Therefore, the peculiar antisymmetric MR evidently stems from the inhomogeneous magnetization process across two Hall junctions, which is induced by the gradient magnetic field.

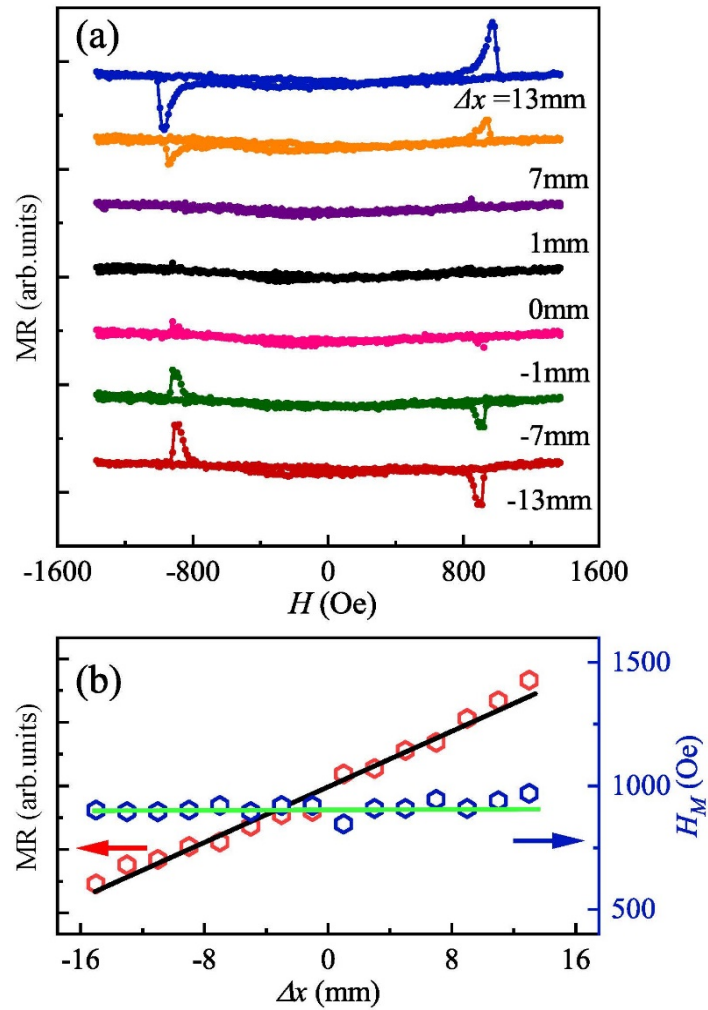

Figure 2. (a) Antisymmetric MR-$H$ curves measured across electrodes A and C at different offset position $\Delta x$. (b) The averaged MR and peak field $H_M$ as a function of $\Delta x$.

To ascertain the effect of magnetization inhomogeneities on the antisymmetric MR, we imposed various magnetic field gradients on the sample. The Hall bar was delicately relocated at different offset position along $x$ axis, and the MR across electrodes A and C was measured accordingly, as shown in Fig. 2(a). The negative MR peak of the ascending field branch is gradually reduced to zero when $\Delta x$ changes from −13 mm to 0, and subsequently reverses its sign as $\Delta x$ changing from 0 to 13 mm. Meanwhile, the positive MR peak of the descending field branch shows a similar evolutionary trend but of opposite sign. It is easily seen that the MR measured under uniform field ($\Delta x = 0$) is kept nearly zero, in consistent with the previous result (Fig. 1(b)). The MR polarity reversal with respect to the offset position $\Delta x$ obviously arises from the opposite spatial distribution of inhomogeneous gradient field along $x$-axis. The values of averaged magnetoresistance, $\mathrm{MR}_A = \frac{\mathrm{MR}(H)-\mathrm{MR}(-H)}{2}$, and the peak field at which the MR reaches its maximum of the ascending field branch, $H_M$, have been derived from Fig. 2(a) to plot as a function of $\Delta x$, as shown in Fig. 2(b). While the $\mathrm{MR}_A$ increases linearly with $\Delta x$, $H_M$ almost remains unchanged. The enhanced magnetic field gradient apparently leads to larger magnetoresistance. Thus, these results definitely demonstrate that the antisymmetric MR is triggered by the inhomogeneous magnetization reversals across the Hall bar, which can be effectively regulated by varying the inhomogeneous field gradient.

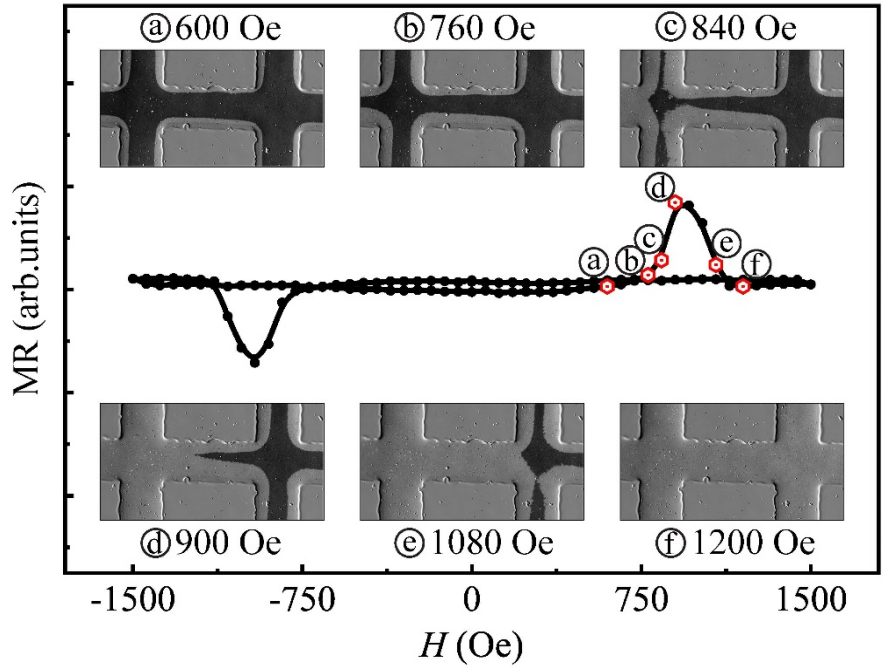

Figure 3. Field dependence of antisymmetric MR and simultaneously measured MOKE images of domain pattern at selective/specified fields.

In order to gain insight into the role of inhomogeneous magnetization texture in the generation of antisymmetric MR, we synchronously characterized the magnetic domain structures by means of MOKE microscopy during transport measurements. For the purpose of reproducing the antisymmetric MR, the sample was intentionally displaced away from the uniform field region. Fig. 3 shows the domain images obtained at selected fields in the ascending field branch. It is noted that the cross-shaped and wedge-shaped domain patterns appear in the junctions and central-arm regions of the Hall bar, respectively. Macroscopically, the discrepancy in domain pattern evolution cross the Hall bar is largely related to both the geometrical restrictions on distinct stripe domain structures and their exactly matched interconnection under inhomogeneous field [26,27] . In the descending field branch, the domain pattern evolution exhibits a similar trend with opposite magnetization (see

Supplemental Material [28]). More specifically, the emergence of nonzero MR is closely correlated with the asymmetric distribution of multidomain texture, which is manifested in the following aspects. (i) The overall area of the unreversed cross-shaped domain (black) inside junction AB is generally smaller than that of junction CD, confirming the asynchronized magnetization reversal of two junctions. (ii) In the central arm region, the wedge-shaped stripe domain with tilting domain walls gradually retreats toward the CD junction with increasing field until it disappears. (iii) As MR reaching its maximum around 900 Oe, the completely reversed cross-shaped domain within junction AB contrasts strongly with the mostly unreversed domain within junction CD, while the tip of the wedge-shaped domain shrinks toward the middle of the center arm. Seemingly, both the asynchronized magnetization reversal of two junctions and the asymmetric wedge domain texture in the central arm are associated with the occurrence of antisymmetric MR. However, the strong coupling between interconnected multidomain structures of the arm and junctions severely hampers efforts to define the dominant aspect from the perspective of magnetic domain analysis. Nevertheless, the complex evolution of the multidomain texture suggests that the single wall model is not simply applicable in our experiment [12,13,16,18].

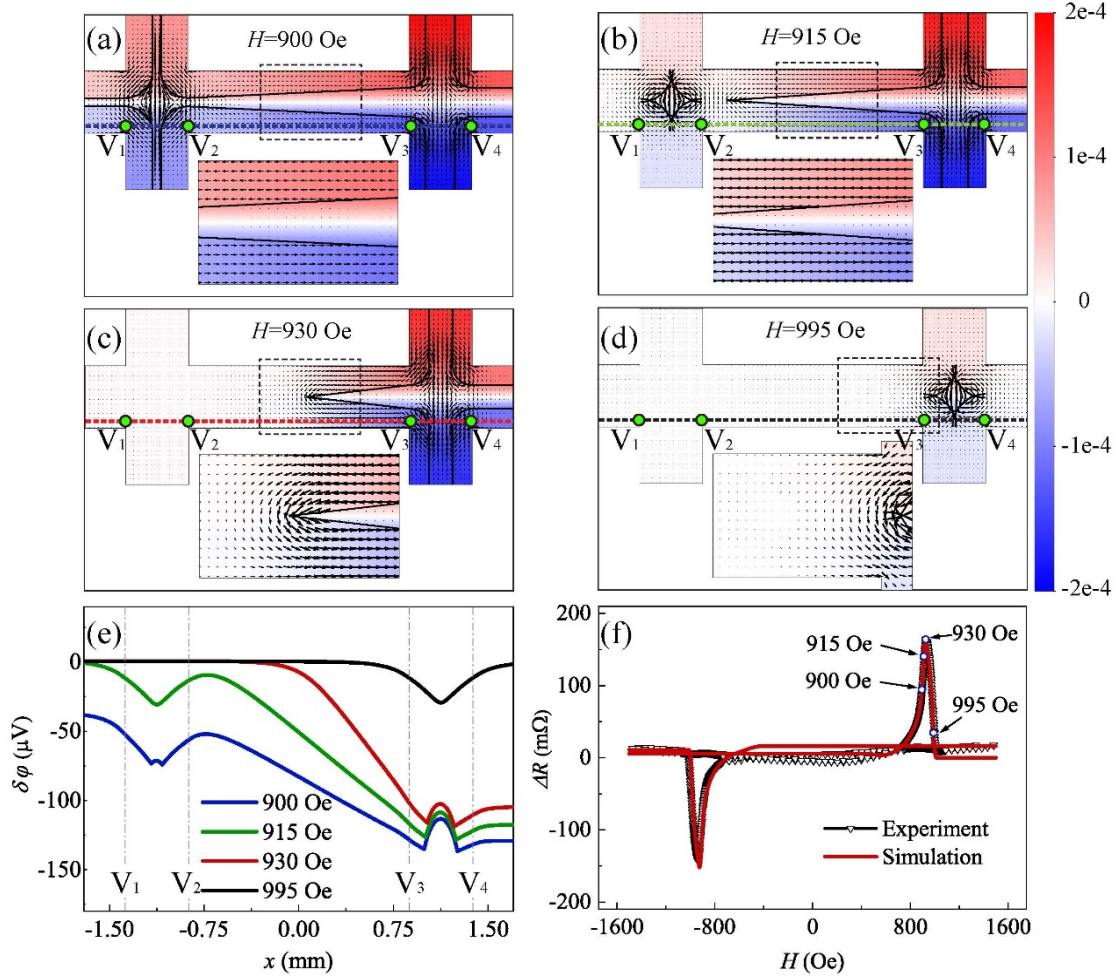

Figure 4. The distributions of $\delta\varphi$ and $\delta\boldsymbol{j}$ of the Hall bar with representative multidomain textures under different fields (a) 900 Oe, (b) 915 Oe, (c) 930 Oe and (d) 995 Oe. The background color (from red to blue in the scale bar of potential) represents the distribution of $\delta\varphi$. The black arrows represent the nonequilibrium currents. The enlarged views of $\delta\boldsymbol{j}$ ($\times 30$) of selected regions are shown in the inset. (e) The distributions of $\delta\varphi$ along lateral edge of the Hall bar at selected fields. The positions of the potential probes are marked as $V_1$, $V_2$, $V_3$ and $V_4$. (f) The experimental and calculated $\Delta R$ as a function of magnetic field.

To comprehend the impact of inhomogeneous multidomain texture on the antisymmetric MR, we calculated the variation of electrical potential $\varphi$ and current density $\boldsymbol{j}$ of the Hall bar with magnetic field. The current density $\boldsymbol{j}$ is determined by $\boldsymbol{E} = \rho_{\uparrow\downarrow}\boldsymbol{j}$, where the resistivity tensor $\rho_{\uparrow\downarrow} = \begin{pmatrix} \rho_0 & \mp\rho_H \\ \pm\rho_H & \rho_0 \end{pmatrix}$ is determined by the spatial distribution of magnetization which is directly derived from representative experimental domain data (see Figs. 4(a) ~ 4(d)). The electrical potential φ is related to $E$ by $E = -\nabla\varphi$. The extra potential $\delta\varphi$ and nonequilibrium current $\delta j$ caused by multidomain texture can be defined as $\delta\varphi = \varphi - \varphi_0$ and $\delta j = j - j_0$, where $\varphi_0$ and $j_0$ represent the electrical potential and current achieved under saturation magnetization, respectively [12,13,16,18].

Figs. 4(a)-4(d) show the calculated distribution of $\delta\varphi$ and $\delta j$ of the Hall bar with representative multidomain textures using the commercial software COMSOL. The nonequilibrium currents $\delta j$ not only appear around the cross-shaped domain in junction regions, but also exhibit noticeably surrounding the tilting domain walls in the central arm region. The enlarged views show that magnitude of $\delta j$ in the vicinity of tilting domain walls increases with $\alpha$, the angle between tilting domain walls. Moreover, the variations of $\delta\varphi$ along the lateral edge of the Hall bar have been examined using four probes ($V_1$ through $V_4$), see Fig. 4(e). Comparing the distributions of $\delta\varphi$ and $\delta j$ between different multidomain textures, we can establish the following results. (i) The substantial drop of $\delta\varphi$ occurs between probes $V_2$ and $V_3$, indicating that the antisymmetric MR is mainly determined by the variation of $\delta\varphi$ in the central arm region. (ii) The $\delta\varphi$ changes significantly in the vicinity of titling domain walls, revealing the dominant role of tilting domain walls in the formation of antisymmetric MR. (iii) With increasing magnetic field, the slope of $\delta\varphi - x$ curve increases, in line with the enhancement of MR (Fig. 4(f)). Therefore, the variations of $\delta\varphi$ and $\delta j$ affected by the domain wall tilting turn out to be crucial for the occurrence of antisymmetric MR, which contrast strongly against the special geometries for experimental observation of the anomalous MR and the single wall

model [12-14,17,18].

To properly describe the effect of domain wall tilting on nonequilibrium current, we derive the following equation by analytic and numerical calculations (see the Supplemental Material [28] for detailed calculations),

$$\nabla \times \delta\boldsymbol{j} = -\frac{j_0 R_s}{\rho_0}\frac{\partial M_z}{\partial x}\hat{\boldsymbol{k}} \qquad (1)$$

where $R_s$ is the Hall coefficient, $Mz$ and $\hat{\boldsymbol{k}}$ are the magnetization and unit vector along the normal direction, respectively. Eq. (1) reveals that any variations of perpendicular magnetization along the current direction, i.e., tiling domain walls, will directly lead to the emergence of nonequilibrium current $\delta j$. More specifically, the $\delta j$ associated with tilting walls can be deduced as $\delta j \approx 2j_0(\rho_H/\rho_0)\tan\frac{\alpha}{2}$ (see Supplemental Material [28]), where $\alpha$ is the angle between the tilting domain walls. It is obvious that the geometry factor $\tan\frac{\alpha}{2}$ is a key parameter for the generation of nonequilibrium currents, which is eventually determined by sample geometry, magnetic anisotropy of films, and the magnetic field gradient in present study. Based on the derived distribution of $\delta\varphi$ and $\delta\boldsymbol{j}$ (Figs. 4(a)-(e)), we further computed the MR between electrodes A and B of the Hall bar. As shown in Fig. 4(f), the calculated $\Delta R$ agrees well with experimental results. Thus, our results evidently indicate that the formation of tilting domain wall is essential to trigger the antisymmetric MR irrespective of whether the sample contains single or multiply domain walls. Furthermore, it is important to emphasize that the anomalous MR can be efficiently realized in magnetic films by delicately controlling the geometry factor of domain wall texture.

In conclusion, our experiments clearly demonstrate that the inhomogeneous magnetization driven by magnetic field gradient gives rise to highly tunable antisymmetric MR in CoTb films. Synchronous measurements of the domain structure and transport properties reveal that such anomalous MR mainly arises from the nonequilibrium current in the vicinity of the tilting domain walls of the multidomain structure. Furthermore, our theoretical calculation and analysis show that the induced nonequilibrium current is basically determined by the geometry factor of the tilting wall texture. The delicate control of MR by properly modulating the geometric parameter demonstrated here can be extended to a broad class of ferromagnetic materials, and provide a new insight into design of novel spin-logic devices.

**Acknowledgements**

This work was supported by the National Key Basic Research Project of China (No.2016YFA0300600), the Strategic Priority Research Program of the Chinese Academy of Sciences (No. XDB33020300), and the National Natural Science Foundation of China (No. 11604375, No.11874416).